\documentclass[slac_one]{revtex4}
\usepackage{graphicx}
\usepackage{fancyhdr}

\usepackage{color}

\newcommand{\epem}   {\ensuremath{\mathrm{e^+e^-}}}
\newcommand{\as}     {\ensuremath{\alpha_{\mathrm{s}}}}
\newcommand{\Petra} {\mbox{\rm PETRA}}
\newcommand{\Lep}{\mbox{LEP}}
\newcommand{\zzero}     {\ensuremath{\mathrm{Z^0}}}
\newcommand{\asmz}   {\ensuremath{\alpha_{\mathrm{s}}(M_{\mathrm{Z^0}})}}
\newcommand{\Jade}{\mbox{\rm JADE}}

\newcommand{\Opal}{\mbox{\rm OPAL}}
\newcommand{\Aleph} {\mbox{\rm ALEPH}}

\newcommand{\ytwothree}   {{\ensuremath{y^{\rm D}_{23}}}} 
\newcommand{\bt}     {\ensuremath{B_{\mathrm{T}}}}
\newcommand{\bw}     {\ensuremath{B_{\mathrm{W}}}}
\newcommand{\cp}     {\ensuremath{C}}
\newcommand{\mh}     {\ensuremath{M_{\mathrm{H}}}}
\newcommand{\thr}    {\ensuremath{1-T}}
\def\d{\hbox{d}}
\newcommand{\muR} {\ensuremath{\mu_R}}
\newcommand{\xmu}    {\ensuremath{x_{\mu}}}
\newcommand{\rs}     {\ensuremath{\sqrt{s}}}
\newcommand{\chisqd} {\ensuremath{\chi^2/\mathrm{d.o.f.}}}
\newcommand{\momn}[2] {\mbox{\ensuremath{\langle#1^{#2}\rangle}}}
\newcommand{\sigtot}     {\ensuremath{\sigma_{\mathrm{tot.}}}}
\newcommand{\dd}    {\ensuremath{\mathrm{d}}}
\newcommand{\nlo}{\mbox{NLO}}
\newcommand{\oaa}    {\ensuremath{\mathcal{O}(\alpha_{\mathrm{s}}^2)}}
\newcommand{\resultJOrhs}  {\ensuremath{      0.1262\pm0.0006\stat\pm0.0010\expt\pm0.0007\had\pm0.0064\theo} }    
\newcommand{\stat}   {\ensuremath{\mathrm{(stat.)}}}
\newcommand{\tot}   {\ensuremath{\mathrm{(tot.)}}}
\newcommand{\expt}               {\ensuremath{\mathrm{(exp.)}}}
\newcommand{\had}               {\ensuremath{\mathrm{(had.)}}}
\newcommand{\theo}              {\ensuremath{\mathrm{(theo.)}}}
     
\newcommand{\anul}   {\ensuremath{\alpha_0}}
\newcommand{\anulmI}   {\ensuremath{\alpha_0(\mu_I)}}

\begin{document}

\title{Measurement of event shapes and \boldmath{$\alpha_s$} in \boldmath{$e^+e^-$}} 

%

\author{C. Pahl}
\affiliation{Max-Planck-Institut f\"ur Physik, F\"ohringer Ring 6, D-80805 Munich, Germany\\
        Excellence Cluster Universe, TU M\"unchen, Boltzmannstr. 2, 
        D-85748 Garching, Germany.}
%
\begin{abstract}
We report results on measurements of the strong coupling \as\ from event shape data
          at \Petra\ and \Lep. These include analyses of their distributions employing the recent NLLO
          calculations and analyses of their moments using NLO and power correction predictions.
\end{abstract}
\maketitle
\thispagestyle{fancy}
\section{INTRODUCTION} 
  In an hadronic event 
    the annihilation of the \epem\ pair into an $\zzero/\gamma$ and its subsequent decay into an
    quark pair is followed by gluon radiation and other processes.
    At an energy scale of about 1 GeV 
    hadronisation takes place and the 
  partons transform into hadrons. 
    The process of hadronisation can only be described by Monte 
    Carlo models or -- more recently -- analytically by power correction models.
    The ultimate goal is to measure the strong coupling; i.e. to test its energy evolution as
    described by QCD and to extract a value at some definite energy scale, 
    commonly taken as the mass of the \zzero\ boson. The most recent world average is
    $\asmz=0.1198\pm0.0010$ \cite{bethke06}.

  The construction of the \epem\ collider experiments \Jade\
 and \Opal\ 
 is quite similar and a consistent measurement can be expected by using both.
\Opal\ operated at centre-of-mass energies from 91 to 209~GeV, \Jade\ at lower energy scales from 
  12 to 44~GeV, where the strong coupling and its energy evolution are stronger.
The experiment \Aleph\ was one of the four LEP experiments.

    The studied event shapes are
    thrust $T$ or \thr, C-parameter \cp\ and total jet broadening \bt; these variables are
    sensitive to the whole event.
    Further heavy jet mass \mh, wide jet broadening \bw\ and Durham two-jet flip pa\-ra\-me\-ter 
    \ytwothree;
    these are sensitive to only one suitable chosen hemisphere of the event. For the definitions of the variables
    we refer to \cite{OPALPR404}.

\section{NNLO ANALYSES OF EVENT SHAPE DISTRIBUTIONS}
\begin{figure*}[htb!]   
   \begin{minipage}[]{8.5cm}\vspace{.4cm}

     \includegraphics[width=.84\textwidth]{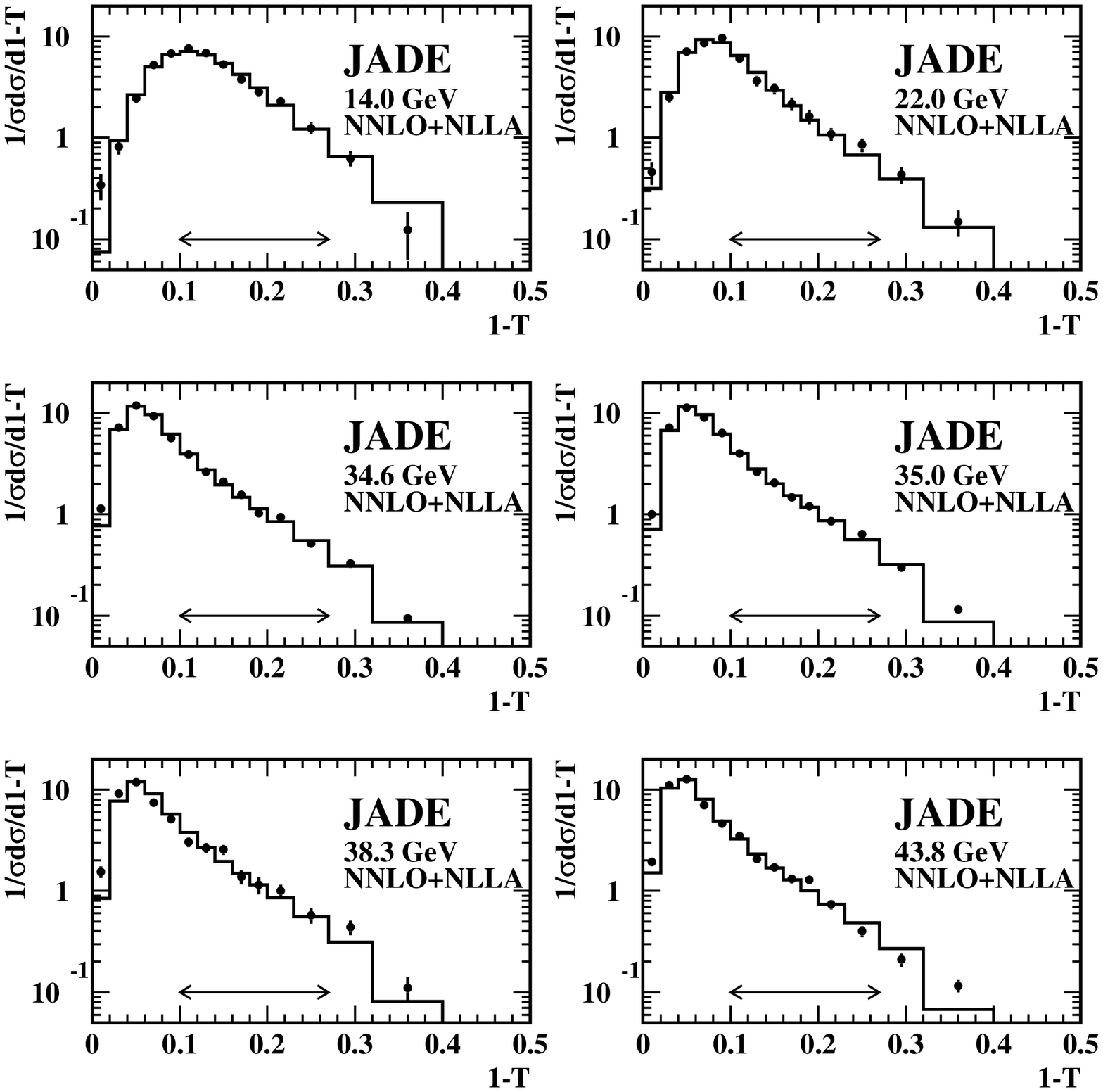} 

   \end{minipage}
   \hspace{.3cm}\begin{minipage}[]{8.5cm}

     \includegraphics[width=.94\textwidth]{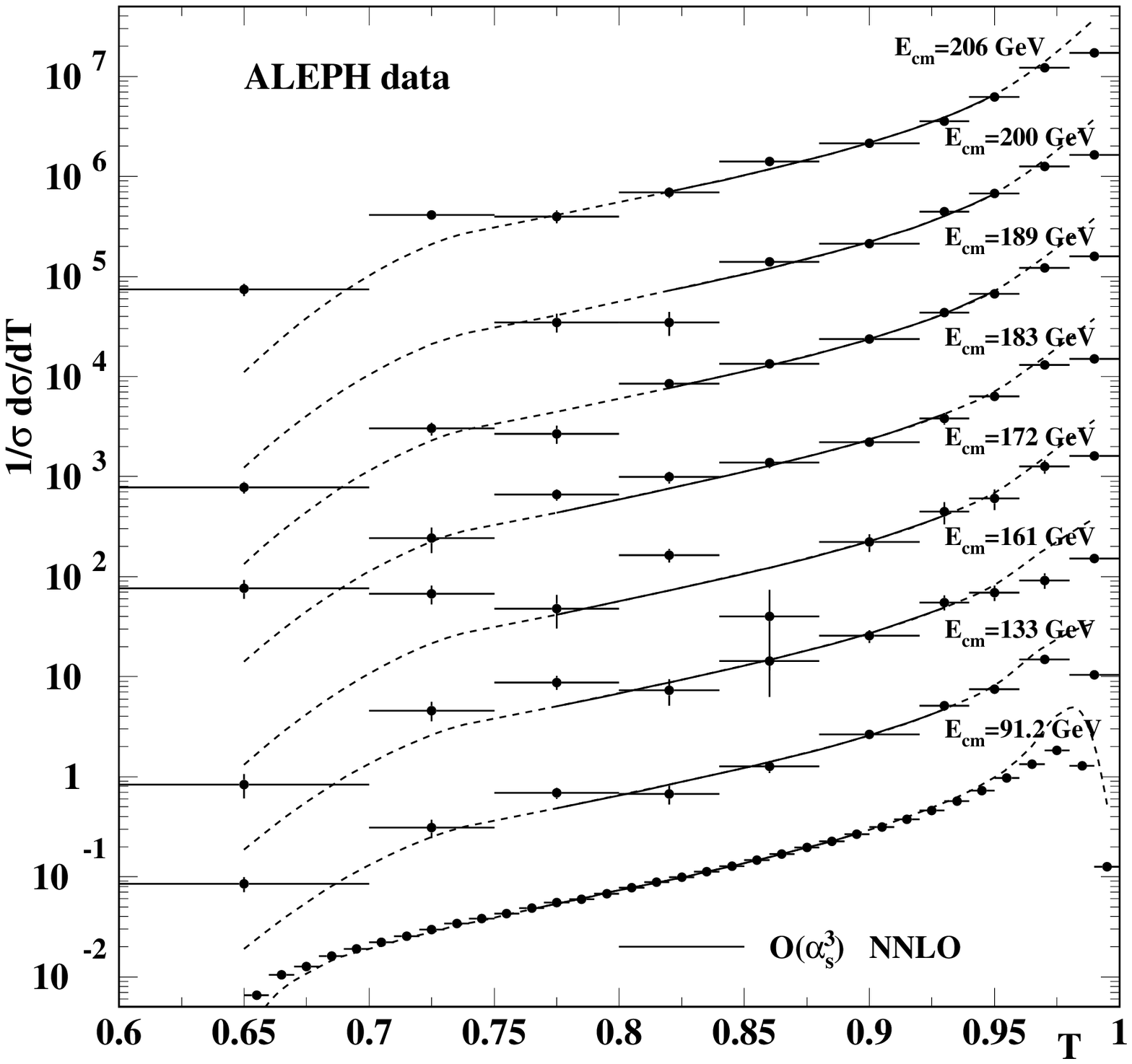}

    \end{minipage}
 \caption{Fits of the NNLO+NLLA prediction to \Jade\ data (prelim.) and
          fits of the NNLO prediction to \Aleph\ data.
          The symbols show the measurement on hadron level with statistical errors. The fit ranges are indicated by 
          arrows rsp. solid lines.\label{fitplots}}
 \end{figure*}

  The recently completed NNLO calculations~\cite{NNLOESs} have been fitted to event shape 
    distributions measured by \Jade~\cite{jadeNNLO} and \Aleph~\cite{asNNLO}. 
  Figure~\ref{fitplots} shows the fits in case
    of thrust; \Jade\ measured $1-T$ at centre-of-mass energies $\rs=$14 to 44\,GeV, while \Aleph\ measured 
    $T$ at 91 to 206\,GeV. In the \Jade\ analysis resummed next-to-next-to-leading logarithms
    (NLLA) are included employing the $\ln R$-matching scheme. In contrast to NLO analyses, these predictions fit 
  well over virtually the whole phase space.

\begin{figure*}[htb!]
    \begin{minipage}[]{7.5cm}
      \hspace{-2.cm} \Large \bf {JADE }{\color{black}\Large (prelim.)}\vspace{-.1cm}

      \hspace{-3.cm}\includegraphics[width=.87\textwidth]{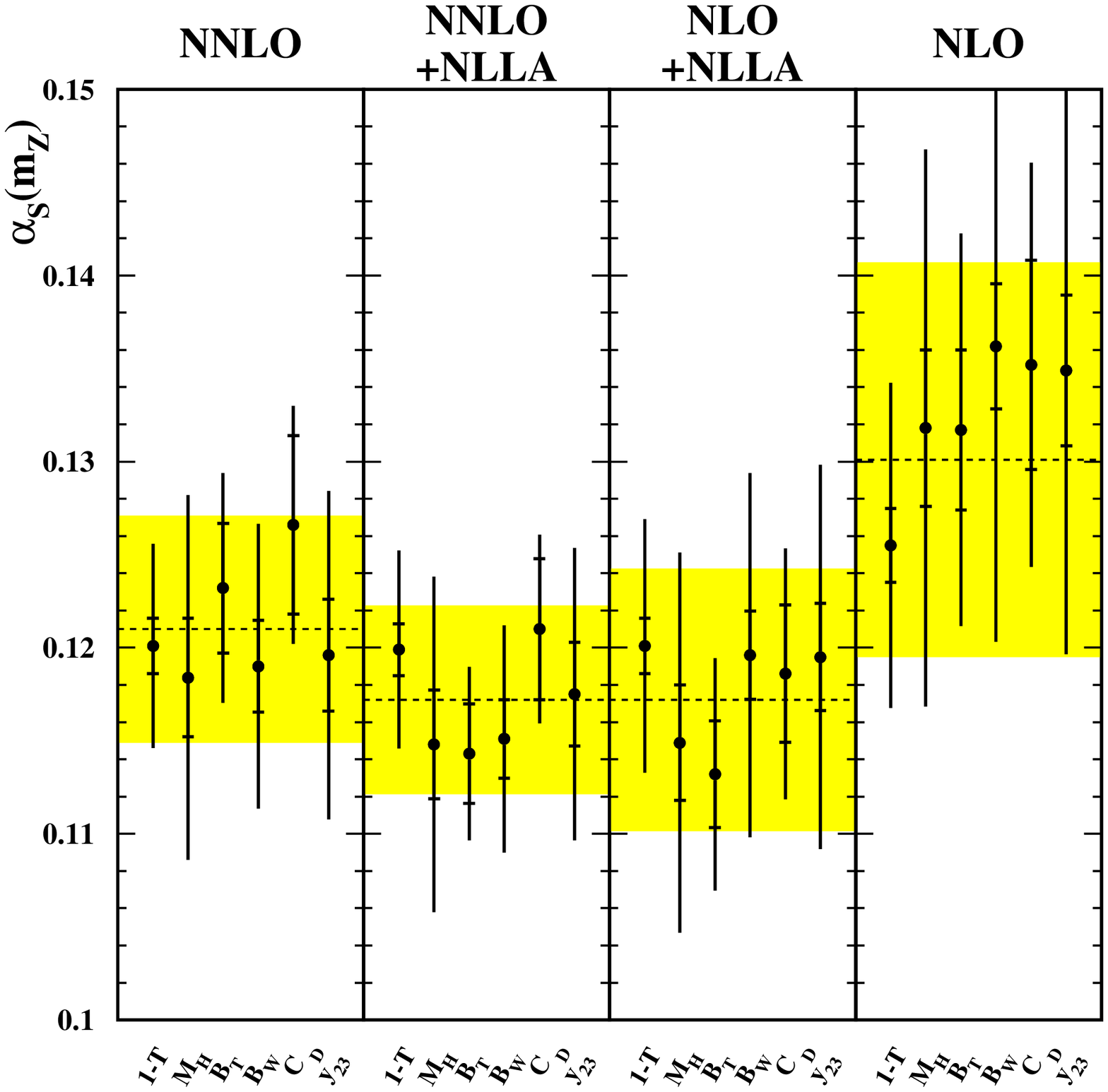} 

  \end{minipage}
  \begin{minipage}[]{7.5cm}
    \vspace{.3cm}

    \hspace{-1.cm}\Large \bf {ALEPH} \vspace{-.3cm}

    \hspace{-2.3cm}\includegraphics[width=1.28\textwidth]{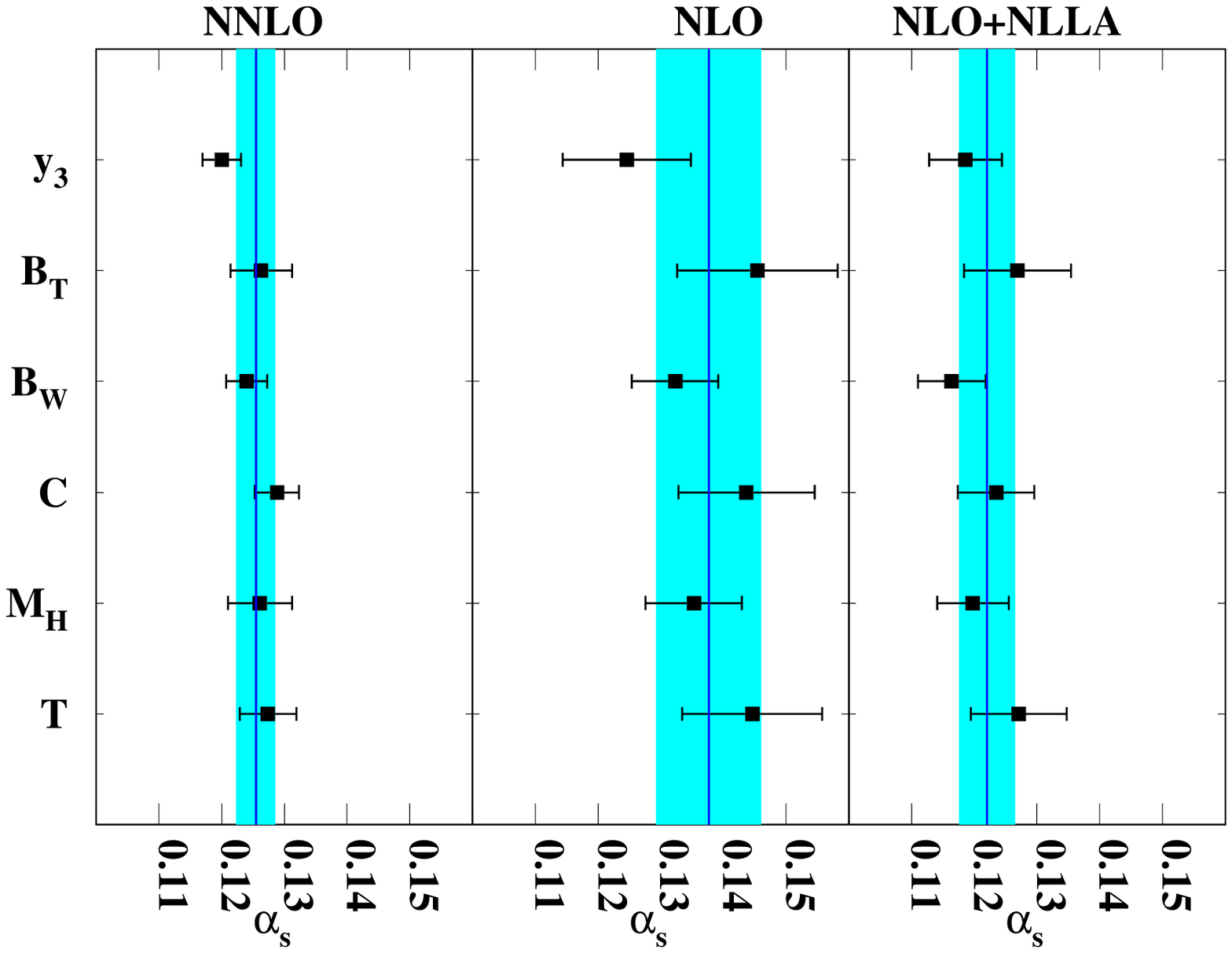}  
  \end{minipage}
  \caption{\asmz\ results from the \Jade\ or \Aleph\ analyses of \thr\ or $T$, \mh, \bt, \bw, \cp, 
    $\ytwothree\equiv y_3$, and combination of
    these variables. The results from the different energy points have been combined. 
    \label{asmzresults}}
  \end{figure*}

  The major uncertainty of NLO analyses~\cite{pedrophd,OPALPR404} comes from varying the renormalisation scale \muR\ 
    for estimating the contribution of missing higher order perturbative terms.
    Therefore the dependency of the fit results on the scale factor 
    $\xmu\equiv\muR/\rs$ has been studied. 
    To compare with the NNLO+NLLA analysis, NNLO, NLO and NLO+NLLA studies have been pursued by \Jade.
    \Aleph\ compares the NNLO analysis with NLO and NLO+NLLA.
    The analyses 
    including NNLO show the smallest dependency on \xmu, which leads to a small 
    scale uncertainty. 
 The resulting \asmz\ values -- combined 
  from the variables thrust, \mh, \bt, \bw, \cp, \ytwothree\ at all analysed energy points
  -- are shown in figure~\ref{asmzresults}.
    The values and their scatter from the different variables
    get smaller by the inclusion of 
the NNLO terms. 
    The same holds for the inclusion of
resummed logarithms.
    The JADE value from the 
    NNLO+NLLA analysis is identical to the value from NLO+NLLA, but scatter and error
    are reduced.
    The theoretical scale uncertainties are considerably reduced by the NNLO terms.
    The \Aleph\ results are similar to the \Jade\ results with somewhat higher central values.
  Combining the \asmz\ values from the NLLO analyses 
  of the different variables, the energy evolution agrees with the QCD prediction of asymptotic freedom. The evolution is 
  stronger in the \Jade\ energy range, but the errors are smaller
    from the \Aleph\ points. The hadronisation effects 
   scale with inverse powers 
    of \rs, so they -- and their errors -- are smaller at high energies.
    Therefore the \Aleph\ result, 
      $\asmz=0.1240\pm0.0008\stat\pm0.0010\expt\pm0.0011\had\pm0.0029\theo=0.1240\pm0.0033$
    is more precise than the \Jade\ result,
      $\asmz=0.1172\pm0.0006\stat\pm0.0020\expt\pm0.0035\had\pm0.0030\theo=0.1172\pm0.0051$.

\section{NLO ANALYSES OF EVENT SHAPE MOMENTS} 
  The $n$-th moment of the distribution of event shape variable $y$ is defined
     as 
     \[
        \momn{y}{n} =\frac{1}{\sigtot} \int y^n
        \frac{\dd\sigma}{\dd y} \dd y \,,
     \]
     The moments have been measured for order $n=1$ to 5 from \Jade~\cite{jadepaper} and \Opal~\cite{OPALPR404} 
   data. 
  With increasing moment order the peak of the integrand is shifted towards the multi jet region,
     so higher moments are sensitive to this region.   

\begin{figure*}[htb!]
  \begin{center}
  \begin{tabular}{c c c}
    \includegraphics[width=.26\textwidth,clip]{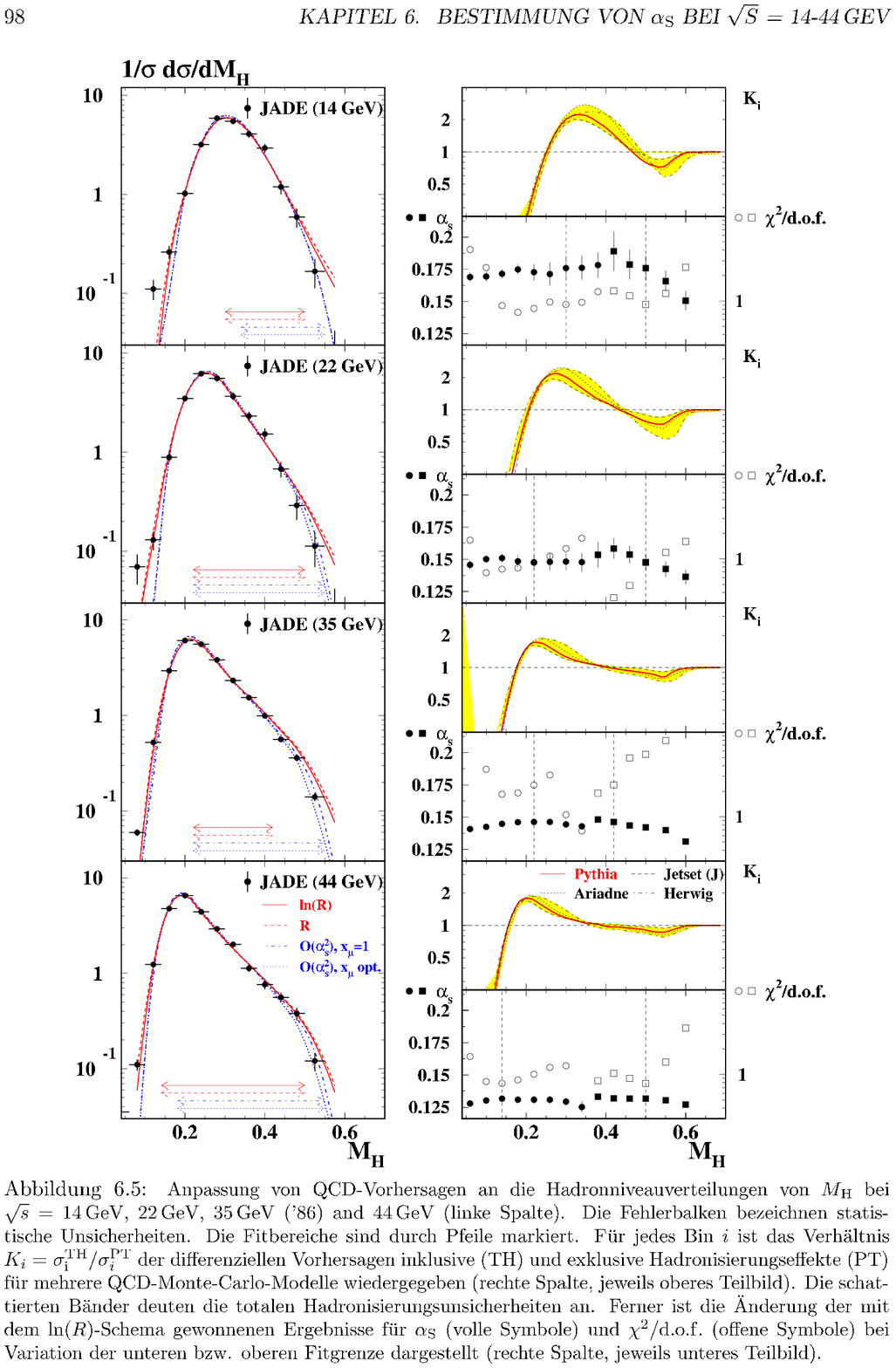} & 
    \includegraphics[width=.26\textwidth,clip]{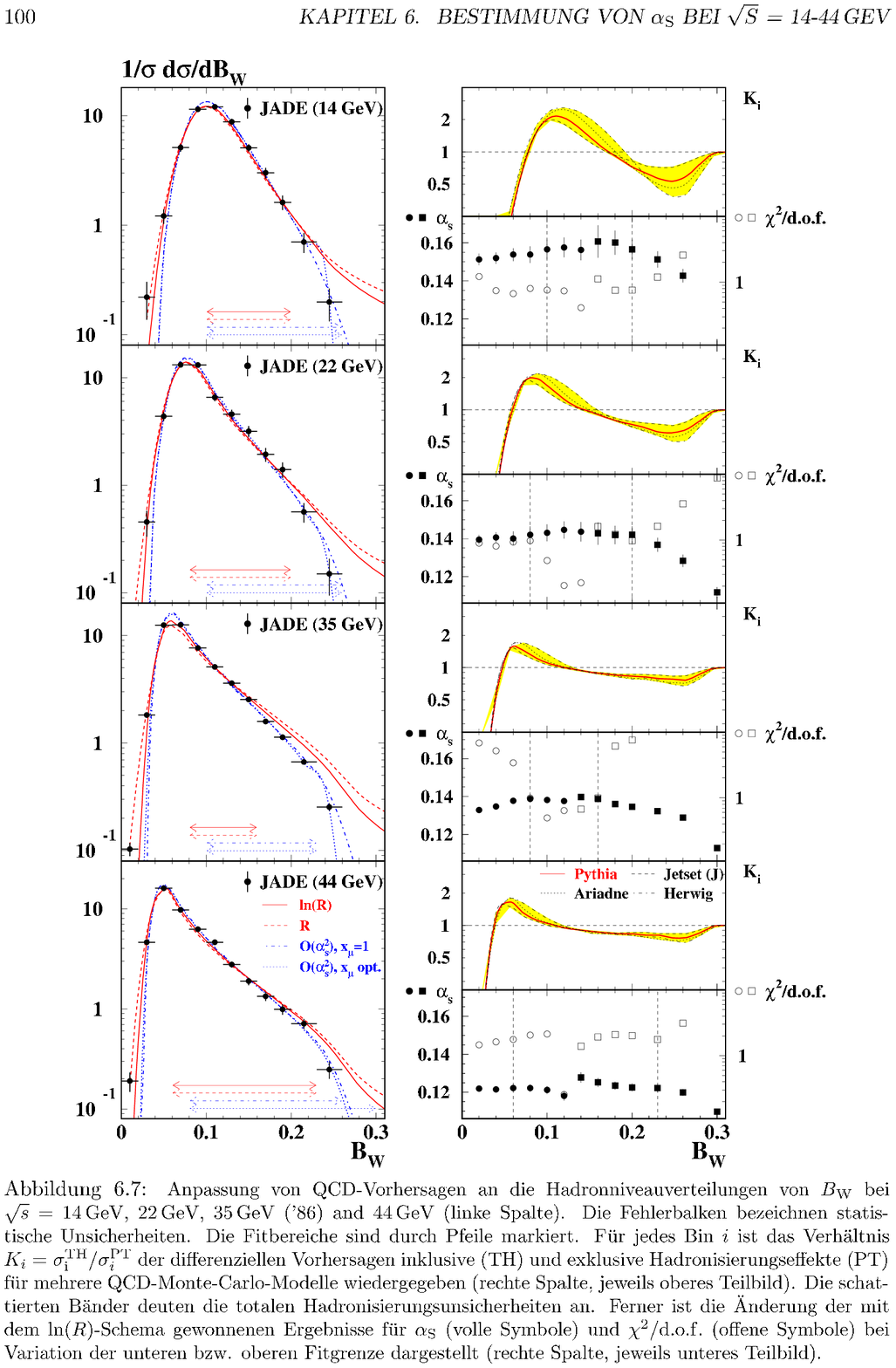} &
    \includegraphics[width=.26\textwidth,clip]{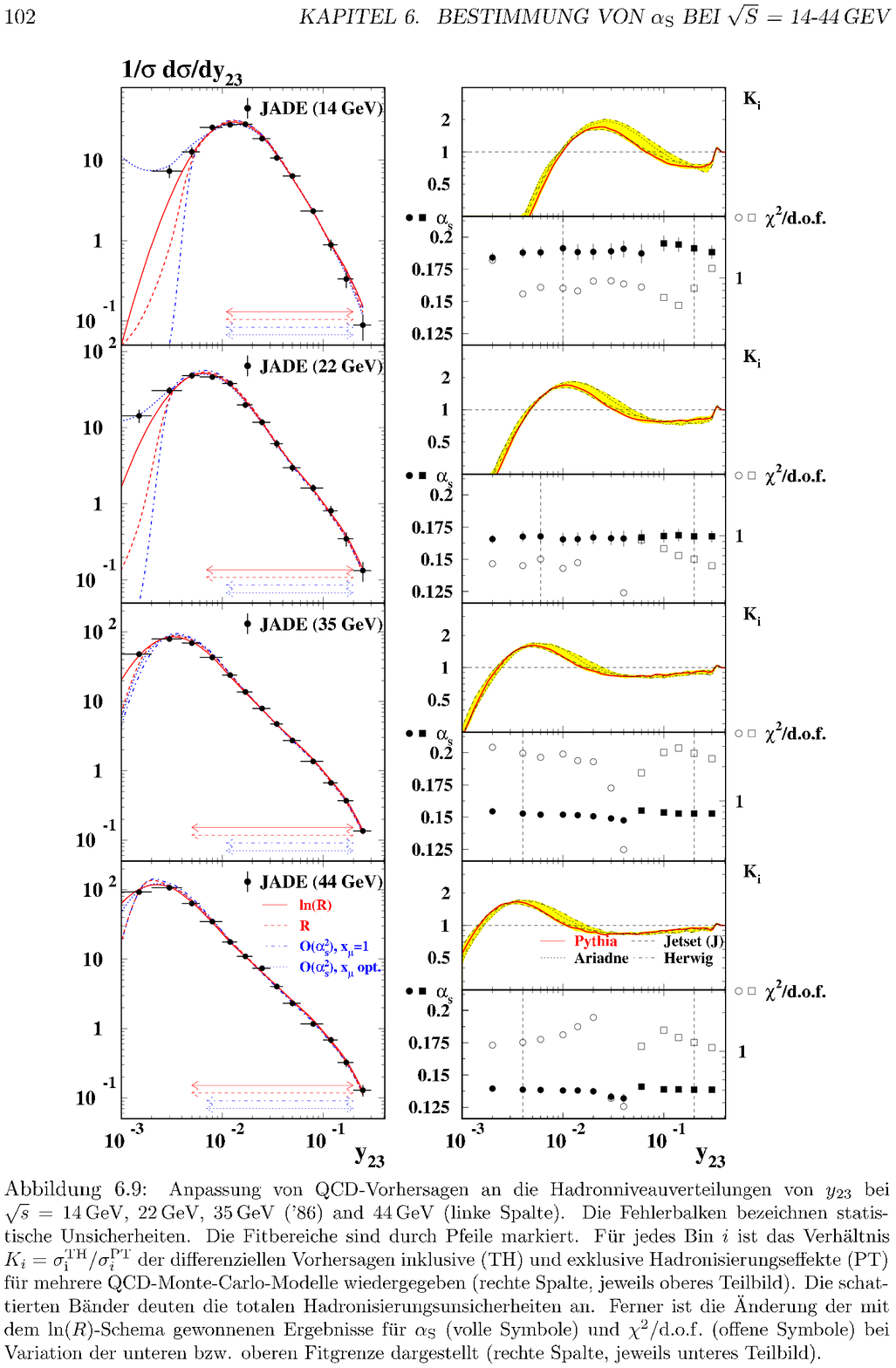} \\
  \end{tabular}
  \end{center}
  \caption{Distributions of the one-hemisphere variables \mh, \bw\ and y$_{23}\equiv\ytwothree$ at 14 GeV; 
           \Jade\ measurement and various fit curves~\cite{pedrophd}.
           As indicated, the dash-dotted curve is to be compared with our analysis. The range where this
           prediction is negative (but not plotted) is marked with an oval starting at zero on the 
           abscissa.\label{pedrosplots}}\vspace{-4.35cm}

\hspace{8.5cm}\hspace{1cm}{\color{blue}\fbox{\begin{minipage}[]{1.3cm}\small compare:\\
                                  ${\cal O}(\as^2)$, $\xmu=1$\end{minipage}}
\vspace{1.cm}\color{black}

\hspace{0cm}{\bf 0\hspace{2.1cm}\mh\hspace{1.2cm}0.7}
\hspace{.2cm}{\bf 0\hspace{1.8cm}\bw\hspace{1.6cm}0.3}
\hspace{.1cm}{\bf $10^{-3}$\hspace{1.3cm}$y_{23}$\hspace{1.5cm}$10^{-1}$}

\vspace{-3.cm}  \setlength{\unitlength}{1cm}
\begin{picture}(30,5){\color{blue}\thicklines
    \put(13.3,4.75){\vector(-3,1){.75}}}      

    \put( 2.2 ,2.6){\color{green}\oval(.65 ,.25)}
    \put( 6.95  ,2.6){\color{green}\oval(.6  ,.25)}
    \put(11.85 ,2.6){\color{green}\oval(.8  ,.25)}
  \end{picture}}
\end{figure*}

 Employing hadronisation corrections from Monte Carlo models the data are 
compared~\cite{jadepaper,OPALPR404} with the \nlo\ prediction 
     $\momn{y}{n}={\cal A}_n \,\as(s) + {\cal B}_n \,\as^2(s)$ (schematically; terms 
     accounting for correct normalisation and scale dependency are suppressed). The \momn{(\thr)}{n}, \momn{\cp}{n} and \momn{\bt}{n} predictions fit the data well. 
     In case of the one-hemisphere observables \momn{\bw}{n}, \momn{(\ytwothree)}{n} and \momn{\mh}{n}, the
     agreement is not good at low energies (but still $\chisqd\simeq10$).   
This effect can be understood from corresponding analyses of
     distributions, see figure~\ref{pedrosplots}.
The \oaa\ coefficient is negative over a large
     part of the two jet range.  So at low energies -- where it
     contributes strongly -- the distribution is unphysically
     negative. As the moment coefficient is 
     the integral over the distribution coefficients, it can be judged as "unphysically
     low" -- i.e. large corrections of higher order can be expected.\footnote{In fact 
     the negative region is smaller in the NNLO prediction \cite{Gehrmann}.}
     Therefore the NLO prediction of \momn{\mh}{1} can not be fitted~\cite{OPALPR404} and the description is
     very bad for \momn{\bw}{1} -- so this fit is not further used.

 The fit results are increasing significantly and systematically with 
     moment order 
     for the two-hemisphere variables. The K-factors ${\cal B}_n/{\cal A}_n$ of the predictions
 qualitatively show
     the same behaviour. So the trend of large higher order contributions for the two-hemisphere
     moments seems to continue; i.e. in the NLO prediction a large part is missing, 
     and this is compensated by an increased \asmz\ value. 
     Combining the results from fitting the more complete predictions (i.e. those satisfying
     ${\cal B}_n < 0.5\cdot{\cal A}_n$) gives 
     $\asmz=\resultJOrhs$, consistent with the world average.

 Hadronisation can also be described by analytical, {\it non perturbative} models. 
     We tested~\cite{CHPphd} several of them with event shape moments data. The {\it dispersive model} 
     by Dokshitzer et al.\ predicts a simple shift of the perturbatively calculated
     distribution when going from parton to hadron level,
       $\d \sigma/\d y = \d \sigma_{\rm part.}/d y(y-a_y \cdot {\cal P}({\alpha_0}))$\,. 
     Here the power correction factor $a_y$ depends on the observable, while ${\cal P}({\alpha_0})$
     does not.
     This description can be integrated giving predictions for the moments. The data fit these predictions 
     well. The high \asmz\ values from the
     two-hemisphere moments are not cured. The parameter \anul\ should be universal -- this is not the case 
     especially for the values from \momn{\bw}{n}, where this problem has been known from a study of the
     distribution~\cite{MovillaFernandez:1998mr}.
     Combining the values from fitting the more complete descriptions gives 
     $\asmz=0.1174\pm0.0050\tot$\,, $\anulmI=0.484\pm0.053\tot\,$.

\section{SUMMARY AND OUTLOOK} 
   The NNLO and NNLO+NLLA fits of event shape distributions measured by JADE and ALEPH show 
   significantly reduced scale uncertainty and reduced scatter from the different variables.
   The predictions fit well over a larger range of the distributions.
   The strong coupling was measured with a precision of 3\% by ALEPH, $\asmz=0.1240\pm0.0033$.
   Moments of event shape distributions have been measured by JADE and OPAL.
   The perturbative NLO prediction is adequate only for some moments; this
   incomplete perturbative description shows up in all studied non perturbative models.
   When passing from first to higher moments, new perturbative and non perturbative
   problems appear.

Better resummation is available: NNNLA has been calculated for thrust in soft collinear effective
     field theory~\cite{BecherSchwartz}. The NNLO predictions for the moments are almost complete~\cite{Gehrmann:privat}. 

{\it Discussion:}
Giulia Zanderighi -- In a recent paper, Weinzierl calculated somewhat different NNLO coefficients than Gehrmann et al.
  Employing these calculations the \as\ values are not expected to change much; but the scatter 
  from the different variables could become still smaller.  
Chris Maxwell -- An analysis of event shape moments within renormalisation group improved perturbation theory would be interesting.
  This way, the DELPHI collaboration was able to describe mean values of event shapes whithout any power corrections.

  \bibliographystyle{unsrt}
  \bibliography{papers}

\begin{thebibliography}{10}

\bibitem{bethke06}
{S. Bethke}.
\newblock Experimental tests of asymptotic freedom.
\newblock {\em Prog. Part. Nucl. Phys.}, 58:351, 2006.

\bibitem{OPALPR404}
{G. Abbiendi et al.}
\newblock Measurement of event shape distributions and moments in e$^+$~e$^-$
  --$>$ hadrons at 91\,{GeV} - 209\,{GeV} and a determination of {\as\,}.
\newblock {\em Eur. Phys. J.}, C40:287, 2005.

\bibitem{NNLOESs}
A.~Gehrmann-De~Ridder, T.~Gehrmann, E.~W.~N. Glover, and G.~Heinrich.
\newblock {NNLO corrections to event shapes in $e^+e^-$ annihilation}.
\newblock {\em JHEP}, 12:094, 2007.

\bibitem{jadeNNLO}
{S. Bethke, St. Kluth, C. Pahl, J. Schieck}.
\newblock {Determination of the strong coupling \as\ from hadronic event shapes
  and NNLO QCD predictions using \Jade\ data}.
\newblock {\em Submitted to Eur. Phys. J.}, C, 2008.

\bibitem{asNNLO}
{G. Dissertori et al.}
\newblock {\em JHEP}, 0802:040, 2008.

\bibitem{pedrophd}
{P. Movilla Fern\'andez}.
\newblock {Studien zur Quantenchromodynamik und Messung der starken
  Kopplungskonstanten \as\ bei \rs=14-44\,GeV mit dem JADE-Detektor}.
\newblock {\em PhD thesis}, 2003.
\newblock {RWTH Aachen}.

\bibitem{jadepaper}
{C. Pahl, S. Bethke, St. Kluth, J. Schieck}.
\newblock {Study of moments of event shapes and a determination of \as\ using
  \epem\ annihilation data from \Jade}.
\newblock {\em Submitted to Eur. Phys. J.}, C, 2008.

\bibitem{Gehrmann}
{T. Gehrmann}.
\newblock Status of {NNLO} jet calculations.
\newblock In {\em Proceedings of {FRIF} workshop on first principles
  non-perturbative {QCD} of hadron jets, {LPTHE}, Paris, France}, 12-14 Jan
  2006.

\bibitem{CHPphd}
{C.~Pahl}.
\newblock {Untersuchung perturbativer und nichtperturbativer Struktur der
  Momente hadronischer Ereignisformvariablen mit den Experimenten JADE und
  OPAL}.
\newblock {\em PhD~thesis, {\tt
  http://nbn-resolving.de/urn:nbn:de:bvb:91-diss-20070906-627360-1-2}}, 2007.
\newblock {TU M\"unchen}.

\bibitem{MovillaFernandez:1998mr}
{P. Movilla Fern\'andez}.
\newblock Event shapes from {JADE} data and studies of power corrections.
\newblock {\em Nucl. Phys. Proc. Suppl.}, 74:384, 1999.

\bibitem{BecherSchwartz}
{T. Becher, M. D. Schwartz}.
\newblock {A precise determination of $\alpha_s$ from LEP thrust data using
  effective field theory}.
\newblock {\em JHEP}, 07:034, 2008.

\bibitem{Gehrmann:privat}
{T. Gehrmann}.
\newblock {\em {Private communication}}.

\end{thebibliography}

\end{document}